\documentclass[12pt]{elsart}
\usepackage{amsmath}
\usepackage{amssymb}
\usepackage{amsfonts}
\usepackage{graphics}
\usepackage{epsfig}

\newcommand{\Pn}{\psi}
\newcommand{\Pb}{\bar{\psi}}

\newcommand{\gv}{\gamma_5}

\newcommand{\la}{\langle}
\newcommand{\ra}{\rangle}
\newcommand{\tr}{\operatorname{Tr}}

\newcommand{\hS}{\hat{S}}
\newcommand{\hpi}{\hat{\pi}}

\newcommand{\FP}{\Delta}

\graphicspath{plots}
\begin{document}
\vspace{-1.0cm}
\begin{flushleft}
{\normalsize DESY 99-098} \hfill\\
{\normalsize HLRZ 99-31} \hfill\\
{\normalsize HUB-EP-99/33} \hfill\\
{\normalsize TPR-99-14} \hfill\\
%\vspace{-0.35cm}
%{\normalsize August 1999} 
\end{flushleft}
\vspace{1.0cm}
\begin{frontmatter}
  \title{\bf Higher--Twist Contribution to Pion \\
    Structure Function: 4--Fermi Operators}

  \author[HH]{S. Capitani},
  \author[RG]{M. G\"ockeler},
  \author[HU]{R. Horsley},
  \author[ZT,FU]{B. Klaus},
  \author[FU]{V. Linke},
  \author[RG]{P.E.L. Rakow}, 
  \author[RG]{A. Sch\"afer} and
  \author[HH,ZT]{G. Schierholz}
  
  \address[HH]{Deutsches Elektronen-Synchrotron DESY, D-22603
    Hamburg, Germany}
  \address[RG]{Institut f\"ur Theoretische Physik, Universit\"at
    Regensburg, D-93040 Regensburg, Germany} 
  \address[HU]{Institut f\"ur Physik, Humboldt-Universit\"at zu Berlin,
    D-10115 Berlin, Germany}
  \address[ZT]{Deutsches Elektronen-Synchrotron DESY,\\
    John von Neumann-Institut f\"ur Computing NIC, D-15738
    Zeuthen, Germany} 
  \address[FU]{Institut f\"ur Theoretische Physik, Freie Universit\"at
    Berlin, D-14195 Berlin, Germany}
  \date{ }  

\begin{abstract}
We present quenched lattice QCD results for the contribution of 
higher-twist operators to the lowest non-trivial moment of the pion
structure function. To be specific, we consider the combination 
$F_2^{\pi^+} + F_2^{\pi^-} - 2 F_2^{\pi^0}$ which has $I = 2$ and 
receives contributions from 4-Fermi operators only. We introduce the
basis of lattice operators. The renormalization of the operators is
done perturbatively in the $\overline{\rm{MS}}$ scheme using the 't
Hooft-Veltman prescription for $\gamma_5$, taking particular care of 
mixing effects. The 
contribution is found to be of $O(f_\pi^2/Q^2)$, relative to the
leading contribution to the moment of $F_2^{\pi^+}$.
\end{abstract}


\end{frontmatter}

\section{Introduction}

The deviations from Bjorken scaling seen in deep-inelastic
structure functions are usually interpreted
as logarithmic scaling violations, as predicted by 
perturbative QCD. There is mounting evidence~\cite{Liuti}, however,
that part of the deviations are due to power
corrections induced by higher-twist effects.  

This is good news: higher-twist operators probe the non-perturbative
features of hadronic bound states beyond the parton model, and they
may provide valuable  
information on the interface between perturbative and non-perturbative
physics, at least in those cases where the operator product expansion 
(OPE) allows a clean separation between short- and long-distance
phenomena. 

The OPE expresses the moments of the structure function as a series of
forward hadron matrix elements of local operators with coefficients
decreasing as powers of $1/Q^2$. In general the expansion takes the
form 
\begin{eqnarray}
\lefteqn{M_n(Q^2) = \int_0^1 \mbox{d}x\,x^{n-2}F_2(x,Q^2)}
\nonumber \\ 
& &= C^{(2)}_n(Q^2/\mu^2,g(\mu))\,A^{(2)}_n(\mu)  
          + \frac{C^{(4)}_n (Q^2/\mu^2,g(\mu))}{Q^2}
          \, A^{(4)}_n (\mu)   
          +  O\Big(\frac{1}{(Q^2)^2}\Big) \nonumber \\
& &\equiv M_n^{(2)}(Q^2) + M_n^{(4)}(Q^2) +
       O\Big(\frac{1}{(Q^2)^2}\Big), 
       \label{eq:ope} 
\end{eqnarray}
where $A(\mu)$ and $C(Q^2/\mu^2, g(\mu))$ are the operator matrix
elements and Wilson coefficients, respectively, renormalized at a
scale $\mu$, with  
the superscript (2 or 4) denoting the twist. Under ideal
circumstances the Wilson coefficients can be calculated perturbatively,
which leaves only the matrix elements to be computed on the
lattice. The leading-twist contribution can be written as
\begin{equation}
  \label{eq:m2}
  M_n^{(2)}=\sum_f Q_f^2 \la x_f^{n-1}\ra,
\end{equation}
where $x_f$ is the energy fraction carried by the quark of flavor
{\small $f$},
and $Q_f$ is the charge of the quark. The higher-twist contributions
usually have no parton model interpretation.

In this paper we will consider the $I=2$ pion structure
function~\cite{Gottlieb:1978,Morelli:1993} 
\begin{equation}
  \label{eq:i2sf}
  F_2^{I=2} \equiv F_2^{\pi^+}+F_2^{\pi^-}-2 F_2^{\pi^0}.
\end{equation}
This belongs to a flavor 27-plet, is
purely higher twist, and receives contributions from 4-Fermi operators
only.\footnote{It thus evades mixing with operators of lower
dimensions. In the general case where we do have mixing, and Wilson
coefficients and higher-twist matrix elements are afflicted with
renormalon ambiguities, calculations are much more difficult. In
particular, one will also have to compute the Wilson coefficients
non-perturbatively. For a first, fully non-perturbative 
calculation of higher-twist contributions to the nucleon structure
function see~\cite{Capitani:1998lm,Capitani:1999}. So far it is,
however, not clear how 4-Fermi operators can be incorporated in such a
calculation.}   
We restrict ourselves to the lowest non-trivial moment $M_2$.
Thus we will have to compute $A_2^{(4)}(\mu)$.
The problem splits
into two separate tasks. The first task obviously is to compute the
pion matrix elements of the appropriate lattice 4-Fermi operators.
The second task is to renormalize these operators at some finite scale
$\mu$. 

The paper is organized as follows. In sect.~2 we identify the
(renormalized) operator whose matrix elements we have to compute 
according to the OPE. In sect.~3 we perform the renormalization of the
4-Fermi operators. This is done to 1-loop order in perturbation theory. In 
sect.~4 we present the results of the lattice calculation. Finally, in
sect.~5 we collect our results and conclude.

\section{4--Fermi Contribution}

The leading twist-2 matrix element $A_2^{(2)}$ that enters $M_2$ is
given by  
\begin{equation}
  \label{eq:ltwist}
  \langle\vec{p}|O_{\{\mu\nu\}}|\vec{p}\rangle =  
 2 A_2^{(2)}[p_{\mu}p_{\nu} - \mbox{trace}] , 
\end{equation}
\vspace{-0.8cm}
\begin{equation}
  \label{eq:ltop}
  O_{\mu\nu} = \frac{\mbox{i}}{2}\Pb\gamma_{\mu} G^2 
  \overleftrightarrow{D}_{\nu} \Pn - \mbox{trace} , 
\end{equation}
where $\{\cdots\}$ means symmetrization, and
$\overleftrightarrow{D}=\overrightarrow{D}-\overleftarrow{D}$. The 
fermion fields are 2-component vectors in flavor space, corresponding to $u$
and $d$ quarks, and the charge matrix is
\begin{equation}
  \label{eq:charge}
  G = \begin{pmatrix}
    e_u & 0 \\
    0 & e_d
  \end{pmatrix} =
  \begin{pmatrix}
    2/3 & 0 \\
    0 & -1/3
  \end{pmatrix} . 
\end{equation}
The states are normalized as
 \begin{equation} 
 \label{minkonorm} 
  \langle\vec{p}\,|\vec{p}\,'\rangle = 
 (2 \pi)^3 \, 2 E_{\vec{p}} \, \delta (\vec{p} -\vec{p}\,') . 
 \end{equation} 
The Wilson coefficient has the form
\begin{equation}
  \label{eq:ltcoeff}
  C_2^{(2)} = 1+O(g^2) .
\end{equation}
This contribution is the same for charged and neutral pions, and so
vanishes when considering the structure function $F_2^{I=2}$. 

The twist-4 matrix element $A_2^{(4)}$ receives, in general,
contributions from a large variety of operators. Here we shall only be
interested in 4-Fermi operators, because these are the only operators
that contribute to $F_2^{I=2}$. Following \cite{Jaffe:1981,Jaffe:1982}
we have
\begin{equation}
  \label{eq:htwist}
  \langle\vec{p}|A^c_{\{\mu\nu\}}|\vec{p}\rangle  = 2
  A_2^{(4)}[p_{\mu}p_{\nu} - \mbox{trace}] , 
\end{equation}
\vspace{-0.8cm}
\begin{equation}
  \label{eq:htop}
  A^c_{\mu\nu} = \Pb G\gamma_{\mu}\gamma_5
  t^a\Pn\Pb G\gamma_{\nu}\gamma_5 t^a\Pn - \mbox{trace}.
\end{equation}
The corresponding Wilson coefficient is given by
 \cite{Jaffe:1981,Jaffe:1982,Shuryak:1982b} 
\begin{equation}
  \label{eq:htcoeff}
  C_2^{(4)} =   g^2(1+O(g^2)).
\end{equation}
The operator (\ref{eq:htop}) is understood to be the renormalized,
continuum operator. 

\section{Perturbative Renormalization}

In the following we will be working in Euclidean space-time. 
The 4-Fermi operators that we need to consider on the lattice are
\begin{equation}
  \label{eq:6ops}
  \begin{split}
    V^c_{\mu\nu}  &= \Pb G\gamma_{\mu} t^a \Pn 
    \, \Pb G\gamma_{\nu} t^a \Pn -\mbox{trace},\\[0.5ex]
    A^c_{\mu\nu}  &= \Pb G\gamma_{\mu} 
    \gamma_5 t^a \Pn \,  \Pb G\gamma_{\nu} \gamma_5 t^a
    \Pn-\mbox{trace}, \\[0.5ex] 
    T^c_{\mu\nu} &= \Pb G\sigma_{\mu\rho} 
    t^a \Pn \,  \Pb G\sigma_{\nu\rho} t^a \Pn-\mbox{trace},\\[0.5ex] 
    V_{\mu\nu} &= \Pb G\gamma_{\mu} \Pn 
    \,  \Pb G\gamma_{\nu} \Pn -\mbox{trace},\\[0.5ex]
    A_{\mu\nu} &= \Pb G\gamma_{\mu} 
    \gamma_5 \Pn \,  \Pb G\gamma_{\nu} \gamma_5 \Pn -\mbox{trace},\\[0.5ex]
    T_{\mu\nu} &= \Pb G\sigma_{\mu\rho} 
    \Pn \,  \Pb G\sigma_{\nu\rho} \Pn-\mbox{trace}.\\ 
  \end{split}
\end{equation}
Summation over repeated indices and symmetrization
in $\mu$, $\nu$ is understood. In our conventions
$\sigma_{\mu\nu}=\frac{\rm i}{2} [\gamma_\mu,\gamma_\nu]$. 
The $I=2$ parts of the operators are related by Fierz transformations:
\begin{equation}
  \label{eq:fierz}
  \begin{split}
    V^c_{\mu\nu} &= -\frac{N_c+2}{4N_c} V_{\mu\nu}   -\frac{1}{4} A_{\mu\nu}
    +\frac{1}{4} T_{\mu\nu}, \\ 
    A^c_{\mu\nu} &= -\frac{1}{4} V_{\mu\nu}  -\frac{N_c+2}{4N_c} A_{\mu\nu}
    -\frac{1}{4} T_{\mu\nu}, \\ 
    T^c_{\mu\nu} &=  \frac{1}{2} V_{\mu\nu}   -\frac{1}{2} A_{\mu\nu}
    -\frac{1}{2N_c} T_{\mu\nu} . \\ 
  \end{split}
\end{equation}
The reason for considering all six operators is that they will
mix under renormalization. In principle, the operators (\ref{eq:6ops})
could also mix with gauge variant, BRST invariant operators. But there
are no such 4-Fermi operators with dimension six, and 2-Fermi
operators do not contribute. 

A 1-loop calculation in the continuum and on the lattice gives
for the respective operator matrix elements
\begin{equation}
  \begin{split}
     \langle p, p'|O_i^{\rm cont}(\mu)| p, p' \rangle 
     &= \sum_j \Big(\delta_{ij} +\frac{g_0^2}{16\pi^2}
     R_{ij}^{\rm cont} 
      \Big) \langle p, p' | O_j^{\rm tree}| p, p' \rangle , \\
     \langle p, p' |O_i^{\rm lat}(a)| p, p' \rangle 
     &= \sum_j \Big(\delta_{ij} +\frac{g_0^2}{16\pi^2} R_{ij}^{\rm lat}
      \Big) \langle p, p' | O_j^{\rm tree}| p, p' \rangle, \\
  \end{split}
\end{equation}
where $|p,p'\rangle$ are quark states in some covariant gauge and
$g_0$ is the bare coupling constant.
The continuum matrix elements are understood to be renormalized in the
$\overline{\rm{MS}}$ scheme at the scale $\mu$, while the lattice matrix
elements are unrenormalized. Note that the tree-level matrix elements
$\langle p, p' | O_j^{\rm tree}| p, p' \rangle$ are the same in both
cases. The lattice and continuum matrix elements are then connected by 
\begin{equation}
\label{eq:contmlatt1}
\begin{split}
\langle p, p' | O_i^{\rm cont}(\mu)| p, p' \rangle = \sum_j \Big( \delta_{ij} 
- &\frac{g_0^2}{16 \pi^2} \big( R_{ij}^{\rm lat}
- R_{ij}^{\rm cont}
\big) \Big) \\[0.7ex]
&\times \,\langle p, p' |O_j^{\rm lat}(a)| p, p' \rangle .\\
\end{split}
\end{equation}
Let us write
\begin{equation}
\label{eq:contmlatt2}
\Delta R_{ij} = R_{ij}^{\rm lat} - R_{ij}^{\rm cont}.
\end{equation}
While $R^{\rm lat}$ and $R^{\rm cont}$ depend on the state, $\Delta R$
is independent of the state and depends only on $a\mu$.

The renormalization constants, that take us from bare lattice to
renormalized continuum numbers, are given by
\begin{equation}
Z_{ij} (a\mu, g) = \delta_{ij} - \frac{g_0^2}{16\pi^2} \Delta R_{ij} .
\end{equation}
We have found it convenient to take $p=p'$. The algebraic manipulations
have been done with the help of FORM, using the 't~Hooft-Veltman prescription
for dealing with the $\gamma_5$ matrices. This is the only prescription that
has proven to give consistent results. Integrating the 't Hooft-Veltman
prescription into FORM was a non-trivial task. The algebraic workload
increased by about an order of magnitude. To check the results, many of the
symbolic calculations were repeated by hand. We follow the method
of~\cite{Kawai} for regularizing the infrared divergences.

\begin{figure}[btp]
  \begin{center}
    \epsfig{file=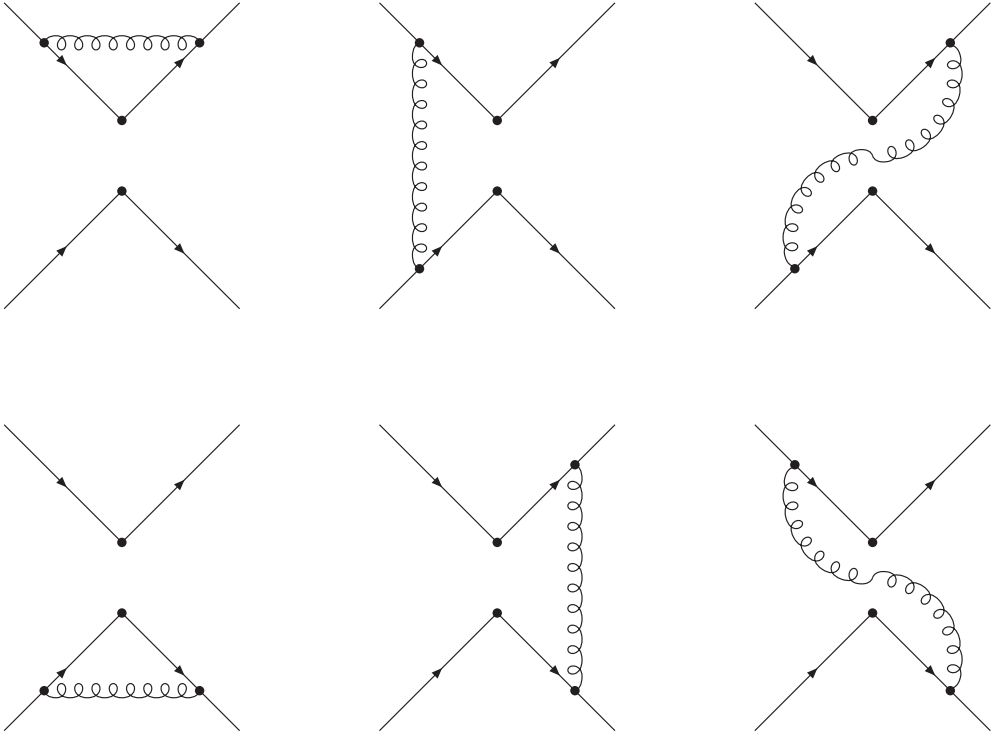, width=1.035\linewidth}
    \caption{The 1-loop diagrams.}
\vspace{0.5cm}
    \label{fig:diagrams}
  \end{center}
\vspace{0.3cm}
\end{figure}

The diagrams that we have calculated are shown in 
fig.~\ref{fig:diagrams}. 
We have not made use of Fierz transformations to reduce the 
number of diagrams, as we could do the numerical integration fast and with
high precision, and also to avoid possible problems with
$d$-dimensional Fierz transformations.

The 1-loop result for the renormalized 4-Fermi operators in the
$\overline{\rm{MS}}$ scheme is
\begin{eqnarray}
\label{eq:1-loop}
    V^c_{\mu\nu}(\mu) &=& V^c_{\mu\nu} -\frac{g_0^2}{16\pi^2} 
    \Big[ \Big( -\frac{1}{2N_c} c_1 + \frac{N_c^2-1}{2N_c} 2S 
    - N_c c_2 \Big) V^c_{\mu\nu} \nonumber \\[0.95ex]
    &+& c_4 \Big(\frac{N_c^2-1}{N_c^2} A_{\mu\nu} +\frac{N_c^2-4}{N_c}
    A^c_{\mu\nu} \Big) 
    - N_c c_3 T^c_{\mu\nu} \Big], \nonumber \\[0.95ex]
    V_{\mu\nu}(\mu) &=& V_{\mu\nu} -\frac{g_0^2}{16\pi^2} 
    \Big[ \frac{N_c^2-1}{2N_c} \Big(c_1 + 2S\Big) V_{\mu\nu} +4 c_4
    A^c_{\mu\nu} \Big], \\[0.95ex]
    A^c_{\mu\nu}(\mu) &=& A^c_{\mu\nu} -\frac{g_0^2}{16\pi^2} 
    \Big[ \Big( -\frac{1}{2N_c} c^{(5)}_1 + \frac{N_c^2-1}{2N_c} 2S 
    - N_c c_2 \Big) A^c_{\mu\nu} \nonumber \\[0.95ex]
    &+& c_4 \Big(\frac{N_c^2-1}{N_c^2} V_{\mu\nu} +\frac{N_c^2-4}{N_c}
    V^c_{\mu\nu} \Big) 
    - c_3 \Big(\frac{N_c^2-1}{N_c^2} T_{\mu\nu} +\frac{N_c^2-4}{N_c}
    T^c_{\mu\nu} \Big)\Big], \nonumber \\[0.95ex] 
    A_{\mu\nu}(\mu) &=& A_{\mu\nu} -\frac{g_0^2}{16\pi^2} 
    \Big[ \frac{N_c^2-1}{2N_c} \Big(c^{(5)}_1 +2S\Big) A_{\mu\nu} +4 c_4
    V^c_{\mu\nu}  
      - 4 c_3 T^c_{\mu\nu} \Big], \nonumber     
\end{eqnarray}
where 
\begin{equation}
S = \log a^2\mu^2 +12.852404 + (1-\alpha)\, (-\log a^2\mu^2 +3.792010)
\end{equation}
is the difference between the (leg) self-energy on the lattice, including 
the tadpole diagram, and in the continuum, and 
\begin{equation}
  \begin{split}
    c_1        &=  -2\log a^2\mu^2 +15.530790 
                   + (1-\alpha)\, (2\log a^2\mu^2 -7.584020), \\[0.5ex]
    c^{(5)}_1  &=  -2\log a^2\mu^2 +5.887758 
                   + (1-\alpha)\, (2\log a^2\mu^2 -7.584020), \\[0.5ex]
    c_2        &=  \frac{1}{2}\log a^2\mu^2 -4.260157 
                   + (1-\alpha)\, (-\log a^2\mu^2 +3.792010), \\[0.5ex]
    c_3        &=    1.205379, \\[0.5ex]
    c_4        &=  -\frac{1}{2}\log a^2\mu^2  +0.094480 . \\
  \end{split}
\end{equation}
Here $\alpha$ is the covariant gauge parameter with $\alpha = 1$ corresponding
to Feynman gauge and $\alpha = 0$ to Landau gauge. The renormalization
constants are gauge invariant. Furthermore, they are independent of $\mu$ and
$\nu$, and hence of the particular representation the operators
reduce to~\cite{rep}. The anomalous dimensions of $V+A$ and $V^c +
A^c$, which are eigenvalues of the mixing matrix, agree with the result found
in~\cite{Okawa}. 

Later on we will be interested in (\ref{eq:1-loop}) for
$N_c=3$ and $\mu = \displaystyle{1/a}$. For this case we have
\begin{equation}
  \label{eq:renormres}
  \begin{split}
    V^c_{\mu\nu}(\mu\!=\!1/a) &= V^c_{\mu\nu} -g_0^2 \big( 0.281578 \,\,
      V^c_{\mu\nu} + 0.000532 \,\, A_{\mu\nu} \\ 
      &+ 0.000997 \,\, A^c_{\mu\nu}  - 0.022899 \,\,
      T^c_{\mu\nu}\big), \\[0.95ex] 
    V_{\mu\nu}(\mu\!=\!1/a) &= V_{\mu\nu} -g_0^2 \big( 0.348170 \,\, V_{\mu\nu} +
      0.002393 \,\, A^c_{\mu\nu} \big), \\[0.95ex] 
    A^c_{\mu\nu}(\mu\!=\!1/a) &= A^c_{\mu\nu} -g_0^2 \big( 0.291756 \,\,
      A^c_{\mu\nu} + 0.000532 \,\, V_{\mu\nu} \\ 
      &+ 0.000997 \,\, V^c_{\mu\nu}  - 0.006785 \,\, T_{\mu\nu}
      - 0.012722 \,\, T^c_{\mu\nu} \big), \\[0.95ex]
    A_{\mu\nu}(\mu\!=\!1/a) &= A_{\mu\nu} -g_0^2 \big( 0.266750 \,\, A_{\mu\nu} 
                + 0.002393 \,\, V^c_{\mu\nu} \\
      & - 0.030533 \,\, T^c_{\mu\nu}
      \big) .\\     
  \end{split}
\end{equation}

Our results will be of use to other applications of 4-Fermi
operators as well. For a 
similar calculation in the context of weak matrix elements, involving
a different set of operators,
see~\cite{Rajan}. 

\section{Lattice Calculation}

\subsection{General Formalism}

To obtain the pion matrix elements of the operators (\ref{eq:6ops}), we need 
to compute the 3- and 2-point functions  
\begin{equation}
  \label{eq:corr}
  \begin{split}
    C^{(3)}_O(t,\tau)&= \frac{1}{V_S}\la{}^S\!\pi^F(t)
    \,O(\tau)\,{}^S\!\pi^{F^\dagger}(0)\ra, \\
    C^{(2)}(t)&= -\frac{1}{V_S}\la{}^S\!\pi^F(t){}^S\!\pi^{F^\dagger}(0)\ra ,\\
  \end{split}
\end{equation}
where $V_S$ is the spatial volume of the lattice. The pion field is
\begin{equation}
  \label{eq:pistat}
    {}^S\!\pi^F(t) = \sum_{x;x_4=t} 
    {}^S\!\Pb_{f\alpha}^a(x )
    F_{ff'}(\gv)_{\alpha\beta}{}^S\!\Pn_{f'\beta}^a(x),
\end{equation}
where $F$ is the flavor matrix. For the different pion states the flavor
matrix takes the form
\begin{equation}
  \label{eq:tau}
  \begin{split}
    \pi^+ &: F = \begin{pmatrix}
      0 &1 \\
      0 &0  
    \end{pmatrix} , \\
    \pi^- &: F = \begin{pmatrix}
      0 &0 \\
      1 &0  
    \end{pmatrix} , \\
    \pi^0 &: F= \frac{1}{\sqrt{2}}\begin{pmatrix}
      1 &0 \\
      0 &-1  
    \end{pmatrix} .\\  
  \end{split}
\end{equation}
The superscript $S$ in (\ref{eq:pistat}) denotes Jacobi-smeared quark fields: 
\begin{equation}
  \label{eq:squark}
  \begin{split}
    {}^S\!\Pn_{f\alpha}^a(x) &= \sum_{y;x_4=y_4=t}
    H^{ab}(x,y;U)\Pn_{f\alpha}^b(y), \\ 
    {}^S\!\Pb_{f\alpha}^a(x) &=
    \sum_{y;x_4=y_4=t}\Pb_{f\alpha}^b(y) H^{ba}(y,x;U), \\    
  \end{split}
\end{equation}
with $H$ being given in \cite{Best:1997qp}, and $U$ denoting the link
variables. 

The 4-Fermi operators in (\ref{eq:6ops}) can be expressed as
\begin{equation}
  \label{eq:gen4f}
  O(\tau) =
  \sum_{x;\,x_4=\tau}\Pb_{f\alpha}^a(x) 
  G_{ff'}\Gamma_{\alpha\beta}^{ab}\, \Pn_{f'\beta}^b(x)
  \Pb_{g\gamma}^c(x)
  G_{gg'}{\Gamma'}_{\gamma \delta}^{\,cd}\,
  \Pn_{g'\delta}^d(x), 
\end{equation}
where $\Gamma$ and
 $\Gamma'$ are products of $\gamma$ and $t^a$ matrices. 

A transfer matrix calculation gives for the time dependence
of the 2-point function
\begin{equation}
  \label{eq:2tdep}
  \begin{split}
    \la{}^S\!\pi^F(t){}^S\!\pi^{F^\dagger}(0)\ra &=
    \tr[\hS^{T-t}\hpi^F\hS^t\hpi^{F^\dagger}] \\
    &=\la 0 |\hpi^{F}|\pi\ra\la\pi|\hpi^{F^\dagger}|0\ra
    e^{-m_\pi t}
    +\la 0 |\hpi^{F^\dagger}|\pi\ra\la\pi|\hpi^{F}|0\ra
    e^{-m_\pi (T-t)},\\
  \end{split}
\end{equation}
where $|\pi\ra$ is the pion ground state, $T$ is the
time extent of the lattice, and $\hS$ is the transfer matrix. All
contributions from excited states have been suppressed.  
The pion states are normalized according to 
 \begin{equation} 
 \label{latnorm} 
 \la \pi(\vec{p}) | \pi(\vec{p}\,') \ra = \delta_{\vec{p}\,\vec{p}\,'} .
 \end{equation} 
The same calculation gives for the 3-point function
\begin{equation}
  \label{eq:3tdep}
  \begin{split}
    \la{}^S\!\pi^F(t)O(\tau){}^S\!\pi^{F^\dagger}(0)\ra &=
    \begin{cases}
      \tr[\hS^{T-t}\hpi^F\hS^{t-\tau}O\hS^\tau\hpi^{F^\dagger}],
      & T\ge t\ge\tau\ge 0, \\
      \tr[\hS^{T-\tau}O\hS^{\tau-t}\hpi^F\hS^t\hpi^{F^\dagger}],
      & T\ge\tau\ge t\ge 0 
    \end{cases} \\[1.5ex]
    &=
    \begin{cases}
      \la 0 |\hpi^F|\pi\ra\la\pi|O|\pi\ra
      \la\pi|\hpi^{F^\dagger}|0\ra
      e^{-m_\pi t}, & T\ge t\ge\tau\ge 0, \\ 
      \la 0
      |\hpi^{F^\dagger}|\pi\ra\la\pi|O|\pi\ra\la\pi|\hpi^F|0\ra
      e^{-m_\pi (T-t)}, & T\ge\tau\ge t\ge 0 .
    \end{cases}\\
  \end{split}
\end{equation}
Thus for the ratio of 3- and 2-point functions we may expect to see a plateau
in $\tau$ at $0\ll\tau\ll t$, and a plateau at $t\ll\tau\ll T$:
\begin{equation}
  \label{eq:ratio}
  R_O(t,\tau) \equiv 
  \frac{C^{(3)}_O(t,\tau)}{C^{(2)}(t)}
  = \begin{cases}
    R_O^{t\geqslant\tau}, & 0\ll\tau\ll t,  \\ 
    R_O^{t\leqslant\tau}, & t\ll\tau\ll T ,
  \end{cases}
\end{equation}
with
 \begin{equation}
  \label{eq:lrplateaux}
  \begin{split}
    R_O^{t\geqslant\tau} &= -
 \la\pi|O|\pi\ra 
      \frac{ e^{-m_\pi t}}
       { e^{-m_\pi t} + e^{-m_\pi (T-t)}},\\
    R_O^{t\leqslant\tau} &= -
 \la\pi|O|\pi\ra 
      \frac{ e^{-m_\pi (T-t)}}
      {e^{-m_\pi t} + e^{-m_\pi (T-t)}}.\\
  \end{split}
\end{equation}
The sum is independent of $t$, and finally we obtain
\begin{equation}
  \label{eq:opval}
% \laO_{\Gamma\Gamma'}\ra \equiv
  \la\pi|O|\pi\ra = -
  R_O^{t\geqslant\tau} 
  - R_O^{t\leqslant\tau}. 
\end{equation}

\subsection{Technical Details}

We will now re-write the 3-point function (\ref{eq:corr}) in terms of
quark propagators. Two different kinds of propagators are emerging, depending
on whether we start from a smeared quark field and end on a local one, or vice
versa. The local-smeared and smeared-local propagators, ${}^{SL}\!\FP$ and
${}^{LS}\!\FP$, are given by 
\begin{equation}
  \label{eq:fprop}
  \begin{split}
    \la{}^S\!\Pn_{f\alpha}^a (x)\Pb_{f'\beta}^b(y)\ra_{\bar{\psi},\psi} &=
     \delta_{ff'}{}^{SL}\!\FP_{\alpha\beta}^{ab}(x,y) \\
    \la\Pn_{f\alpha}^a (x){}^S\!\Pb_{f'\beta}^b(y)\ra_{\bar{\psi},\psi} &=
     \delta_{ff'}{}^{LS}\!\FP_{\alpha\beta}^{ab}(x,y) ,
  \end{split}
\end{equation}
with 
\begin{equation}
  \label{eq:proprel}
  {}^{LS}\!\FP(x,y) =\gv{}^{SL}\!\FP(y,x)^\dagger\gv .
\end{equation}
(We have suppressed the dependence on $U$.)

For $I=2$, and after having summed over flavor indices, we obtain
\begin{eqnarray}
   C^{(3)}_O(t,\tau) &=& (e_u - e_d)^2 \frac{1}{V_S} 
  \sum_{x;\,x_4=t}\sum_{y;\,y_4=\tau}\sum_{z;\,z_4=0}  
    \big\{  \nonumber \\ 
 & &\la\,\tr[{}^{LS}\!\FP(y,z)\gv{}^{SL}\!\FP(z,y)\Gamma'
    {}^{LS}\!\FP(y,x)\gv{}^{SL}\!\FP(x,y)\Gamma]\,\ra_U 
 \nonumber \\ 
    &+& \la\,\tr[{}^{LS}\!\FP(y,z)\gv{}^{SL}\!\FP(z,y)\Gamma
    {}^{LS}\!\FP(y,x)\gv{}^{SL}\!\FP(x,y)\Gamma']\,\ra_U     
  \label{eq:gen4f2} \\
    &-&\la\,\tr[{}^{LS}\!\FP(y,z)\gv{}^{SL}\!\FP(z,y)\Gamma']
    \tr[{}^{LS}\!\FP(y,x)\gv{}^{SL}\!\FP(x,y)\Gamma]\,\ra_U  
 \nonumber \\ 
    &-&\la\,\tr[{}^{LS}\!\FP(y,z)\gv{}^{SL}\!\FP(z,y)\Gamma]
    \tr[{}^{LS}\!\FP(y,x)\gv{}^{SL}\!\FP(x,y)\Gamma']\,\ra_U
 \big\} . \nonumber    
\end{eqnarray}
One of the spatial sums can be eliminated by making use of translational
invariance, finally giving
\begin{eqnarray}
   C^{(3)}_O(t,\tau) &=& (e_u - e_d)^2  
  \sum_{x;\,x_4=t-\tau}\sum_{z;\,z_4=-\tau}  
    \big\{  \nonumber \\ 
 & &\la\,\tr[{}^{LS}\!\FP(0,z)\gv{}^{SL}\!\FP(z,0)\Gamma'
    {}^{LS}\!\FP(0,x)\gv{}^{SL}\!\FP(x,0)\Gamma]\,\ra_U 
 \nonumber \\ 
    &+& \la\,\tr[{}^{LS}\!\FP(0,z)\gv{}^{SL}\!\FP(z,0)\Gamma
    {}^{LS}\!\FP(0,x)\gv{}^{SL}\!\FP(x,0)\Gamma']\,\ra_U     
  \\ 
    &-&\la\,\tr[{}^{LS}\!\FP(0,z)\gv{}^{SL}\!\FP(z,0)\Gamma']
    \tr[{}^{LS}\!\FP(0,x)\gv{}^{SL}\!\FP(x,0)\Gamma]\,\ra_U  
 \nonumber \\ 
    &-&\la\,\tr[{}^{LS}\!\FP(0,z)\gv{}^{SL}\!\FP(z,0)\Gamma]
    \tr[{}^{LS}\!\FP(0,x)\gv{}^{SL}\!\FP(x,0)\Gamma']\,\ra_U
 \big\} .   \nonumber  
\end{eqnarray}

All 3-point functions can be built from 
\begin{equation}
  \label{eq:Q}
  Q_{\alpha\beta}^{ab}(t)=\sum_{x;\,x_4=t}
  {}^{LS}\!\FP_{\alpha\gamma}^{ac}(0,x)(\gv)_{\gamma\delta}
  {}^{SL}\!\FP_{\delta\beta}^{cb}(x,0) ,
\end{equation}
where the first propagator can be obtained from the second one by means of
(\ref{eq:proprel}). 
In terms of (\ref{eq:Q}) the 3-point function reads
\begin{equation}
\label{eq:qform}
\begin{split}
   C^{(3)}_O(t,\tau) &= (e_u - e_d)^2 \big\{  
 \langle \,\tr[ Q(-\tau) \Gamma' Q(t-\tau) \Gamma ] \,\rangle_U \\
&+\langle \tr[ Q(-\tau) \Gamma Q(t-\tau) \Gamma'] \,\rangle_U \\
&-\langle\, \tr[ Q(-\tau) \Gamma'] \tr[ Q(t-\tau) \Gamma ]
 \,\rangle_U \\ 
&-\langle\, \tr[ Q(-\tau) \Gamma ] \tr[ Q(t-\tau) \Gamma']
 \,\rangle_U 
 \big\} . \\
\end{split}
\end{equation} 
\begin{figure}[htbp]
  \begin{center}
    \epsfig{file=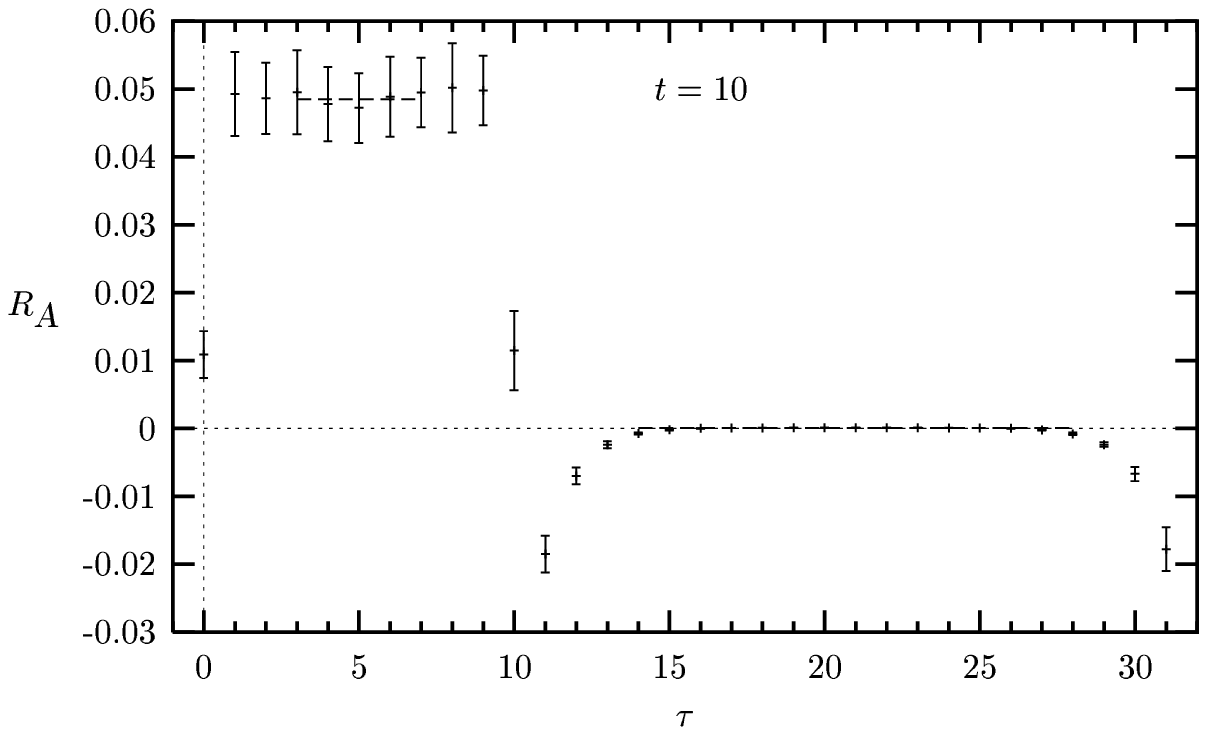, width=.85\linewidth}
\vspace*{0.3cm}

    \epsfig{file=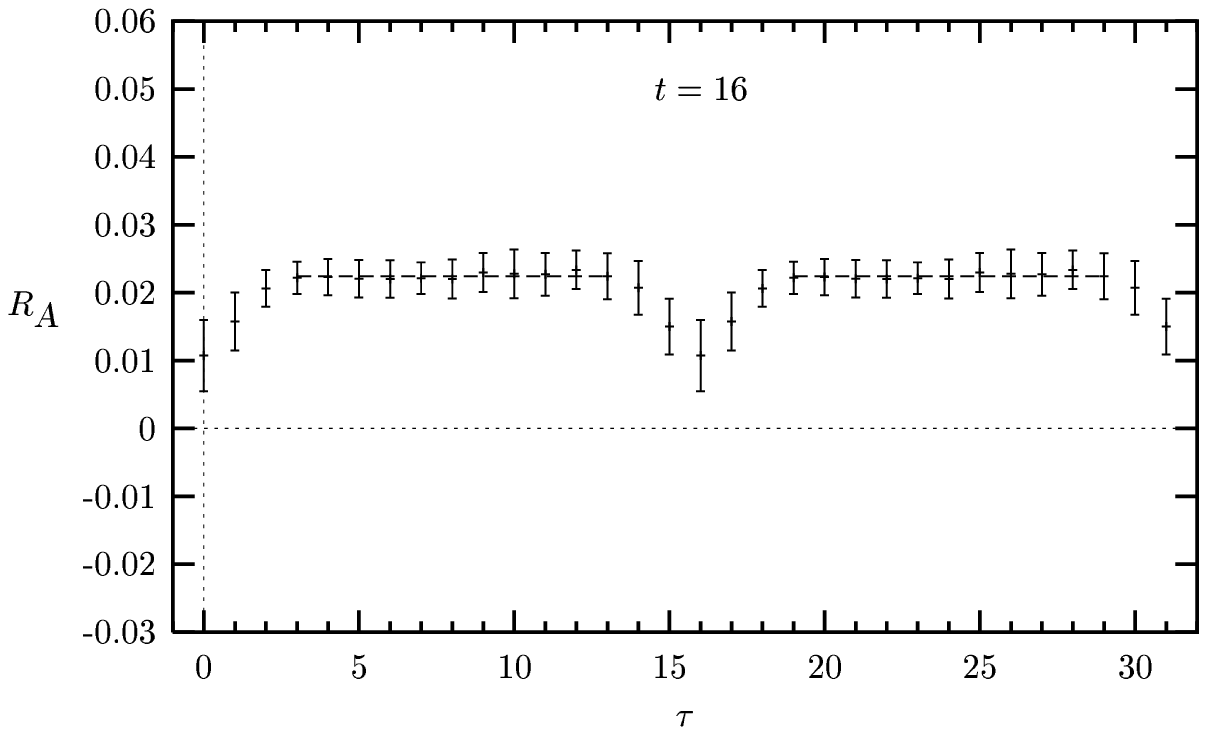, width=.85\linewidth}
\vspace*{0.3cm}

    \epsfig{file=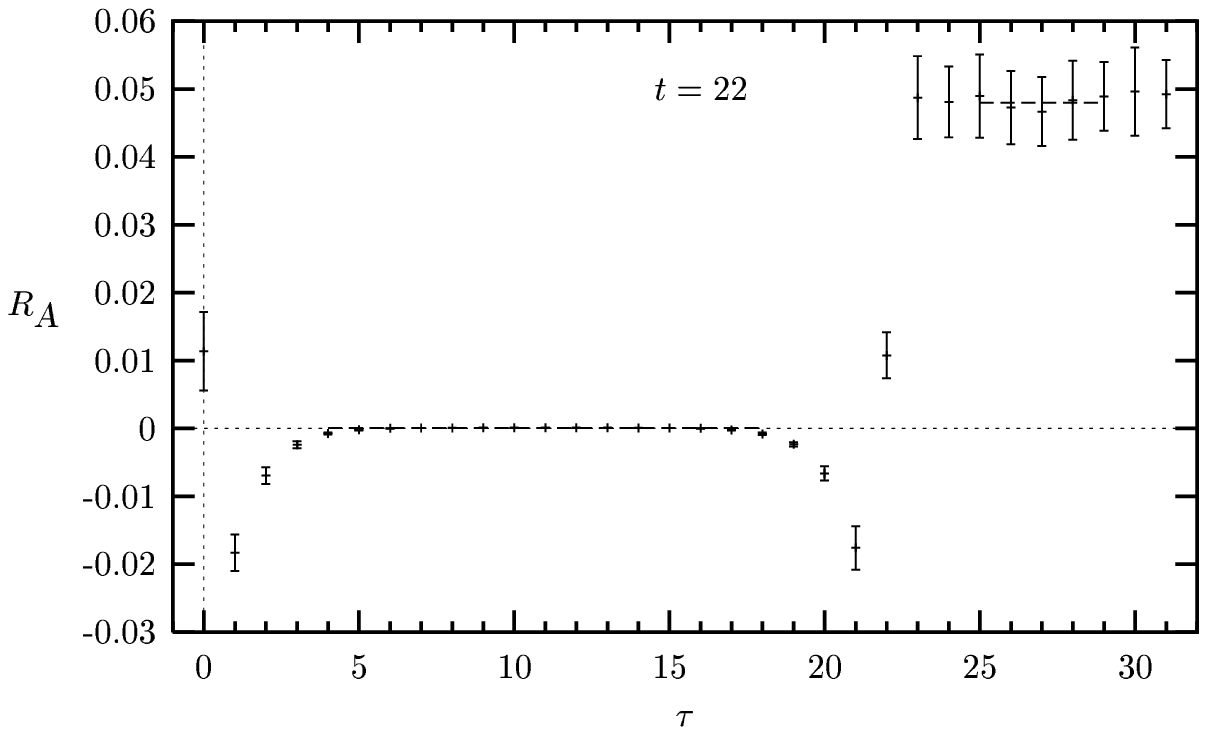, width=.85\linewidth}  
\vspace*{0.3cm}
    \caption{The ratio R for the $44$-component of the operator
      $A$ at $t=10$, 16 and 22 for $\kappa = 0.1515$. The fit
      intervals and the results of the fit 
      are shown by the dashed lines.}
    \label{fig:plat}
  \end{center}
\end{figure}
By computing the propagators from a local source at $t=0$ to all $t<T$
using sink-smearing we obtain the 3-point functions for all $t$ and $\tau$. 

\subsection{Results of the Simulation}

The numerical calculations are done on a $16^3\times 32$ lattice at $\beta
\equiv 6/g_0^2 = 6.0$ in the quenched approximation. 
%We distinguish between
%$g_0^2$ and $g_0^2$, the couplings used in the perturbative and numerical
%calculations, respectively. 
We use Wilson fermions. To be able to extrapolate
our results to the chiral limit, the calculations are done at three values
of the hopping parameter, 
$\kappa = 0.1550$, 0.1530 and 0.1515. This
corresponds to physical quark masses of about 70, 130 and 190 MeV,
respectively. For the gauge update we use a 3-hit Metropolis sweep followed
by 16 overrelaxation sweeps, and we repeat this 50 times to generate a new
configuration. Our data sample consists of 400 configurations.

In the following we shall restrict ourselves to zero momentum pion states and
operators with $\mu = \nu =4$. In fig.~\ref{fig:plat} we show the ratio
(\ref{eq:ratio}) for the operator $A$ at three different values of $t$. We
find very good plateaus in $\tau$. In our final fits we have averaged
over $t$ values around $T/2$. 

To obtain continuum results we have to multiply each quark field by
$\sqrt{2\kappa}$, and a factor of $2m_\pi$ is needed to
convert to the continuum normalization of states:
\begin{equation}
\label{eq:opnorm}
\langle \pi|O|\pi \rangle^{\rm cont} = (2\kappa)^2\, 2m_\pi \, 
\langle \pi|O|\pi \rangle^{\rm lat} .
\end{equation}
For the reduced matrix element $A_2^{(4)}$ (in (\ref{eq:htwist})) we
then find 
\begin{equation}
  \label{eq:a4ar}
 A_2^{(4)} =  \frac{4}{3}\,\frac{(2\kappa)^2}{m_\pi}\,\langle
 \pi|A^c_{44}|\pi \rangle^{\rm lat}, \; \langle \pi|A^c_{44}|\pi
 \rangle^{\rm lat} = - 
 R_{A^c_{44}}^{t\geqslant\tau} - 
 R_{A^c_{44}}^{t\leqslant\tau}, 
\end{equation}
where $A^c_{44}$ is the renormalized operator, as given in
(\ref{eq:1-loop}).

\begin{table}[bhtp]
  \begin{center}
\vspace{0.6cm}
    \begin{tabular}[btp]{|l|l|l|} \hline
      $\kappa$ & $af_{\pi B}$ & $am_\pi$ \\ \hline
      0.1515 & 0.122(2) & 0.5037(8) \\
      0.1530 & 0.113(2) & 0.4237(8) \\
      0.1550 & 0.098(2) & 0.3009(10) \\ \hline
    \end{tabular}
\vspace{0.75cm}
    \caption{Pion masses and unrenormalized (bare) decay constants.}
    \label{tab:fpimpi}
  \end{center}
\vspace{0.3cm}
\end{table}

\begin{table}[hbtp]
\begin{center}
\vspace{0.6cm}
  \begin{tabular}[btp]{|c|l|l|l||l|l|} \hline
\multicolumn{1}{|c}{$\bar{O}$} &
\multicolumn{3}{|c||}{$\kappa$} & 
\multicolumn{2}{c|}{Chiral Limit} \\ \hline
    & $\;0.1515$ & $\;0.1530$ & $\;0.1550$ & $\kappa_c=0.15717$ &
    $\;\;\chi^2$ \\\hline 
    $\bar{A}\,f^{-2}_{\pi B}$ & \;0.561(13) & \;0.546(17) & \;0.514(25) &
    \;\;\;0.490(37) 
    & 0.0593\\
    $\bar{V}\,f^{-2}_{\pi B}$ & -0.139(9) & -0.154(13) & -0.212(23) &
    \;\;-0.237(31) & 0.897\\
    $\bar{T}\,f^{-2}_{\pi B}$ & -0.207(21) & -0.200(30) & -0.197(47) &
    \;\;-0.190(67) & 0.00397\\
    $\bar{A}^c\,f^{-2}_{\pi B}$ & -0.147(11) & -0.139(16) & -0.111(27) &
    \;\;-0.098(38) & 0.111\\
    $\bar{V}^c\,f^{-2}_{\pi B}$ & -0.134(12) & -0.122(17) & -0.089(29) &
    \;\;-0.071(40) & 0.113\\
    $\bar{T}^c\,f^{-2}_{\pi B}$ & -0.315(30) & -0.317(43) & -0.330(68) &
    \;\;-0.334(96) &
    0.00880\\ \hline
  \end{tabular} 
\vspace{0.75cm}
  \caption{The unrenormalized, reduced matrix
    elements $\bar{O}$, together with their
    extrapolations to the chiral limit.}
  \label{tab:latres}
\end{center}
\vspace{0.3cm}
\end{table}

The lattice pion masses are given in table~\ref{tab:fpimpi}.
In the following we shall express the dimensionful matrix elements in
terms of the pion decay constant $f_\pi$, whose unrenormalized values
are also given in the table. The pion masses and the decay constants 
are taken from \cite{Gockeler:1998fn}, where we had a slightly higher
statistics.  

\begin{figure}[htbp]
  \begin{center}
\epsfig{file=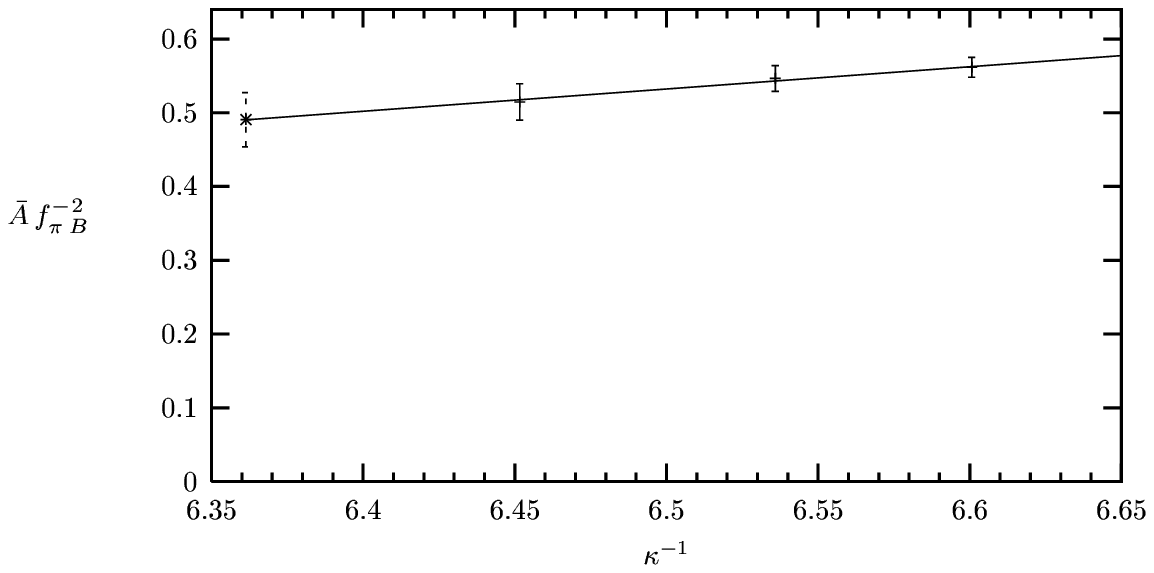, width=.85\linewidth}
\vspace*{0.3cm}

    \epsfig{file=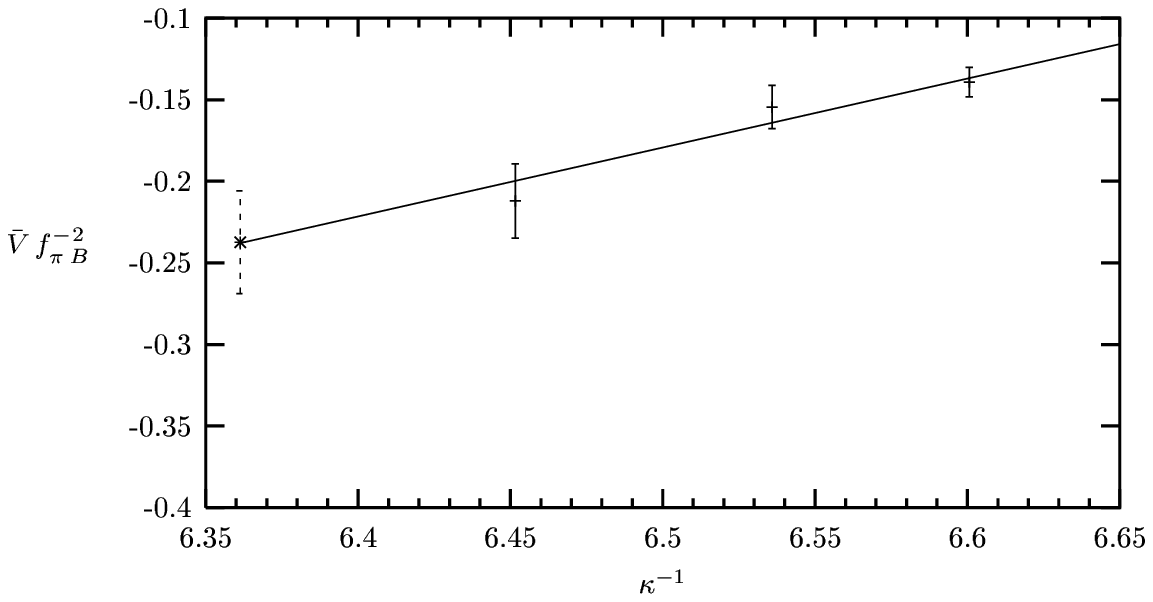, width=.85\linewidth}
\vspace*{0.3cm}

    \epsfig{file=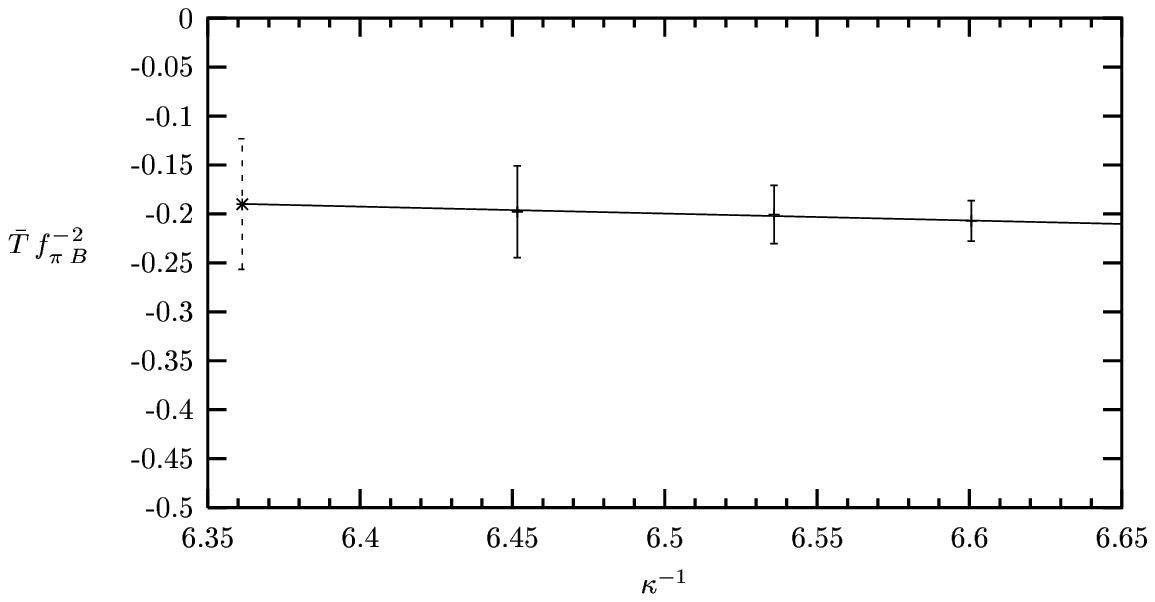, width=.85\linewidth}
\vspace*{0.3cm}
    \caption{The unrenormalized, reduced matrix elements $\bar{A}$,
      $\bar{V}$ and $\bar{T}$, and their extrapolations to the chiral
     limit.} 
    \label{fig:chifit1}
  \end{center}
\vspace{0.4cm}
\end{figure}

\begin{figure}[htbp]
  \begin{center}
    \epsfig{file=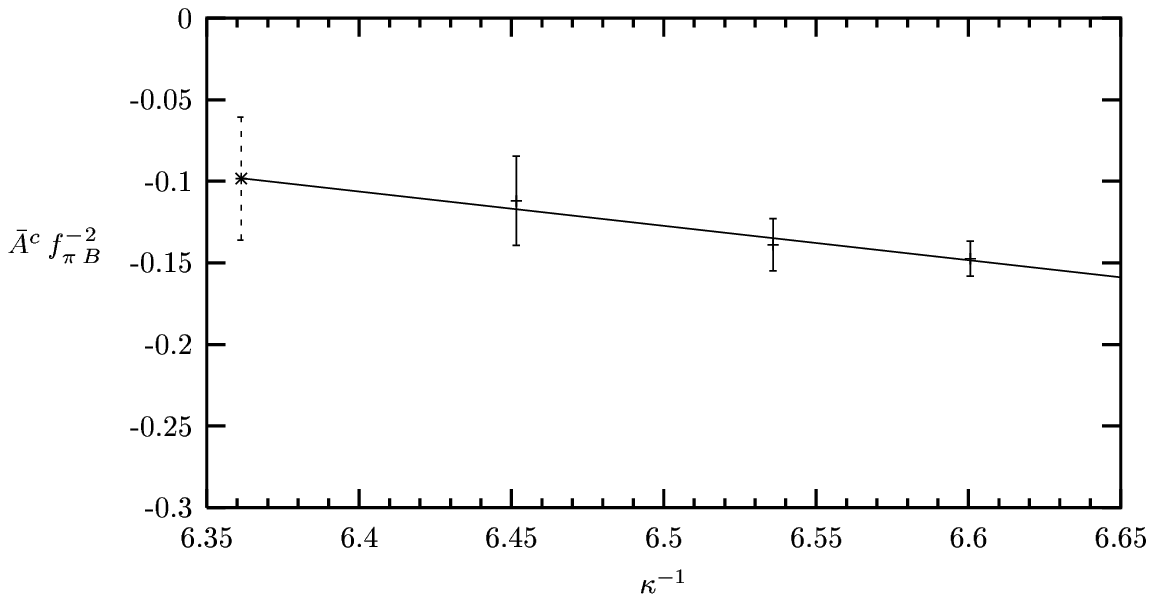, width=.85\linewidth}
\vspace*{0.3cm}

    \epsfig{file=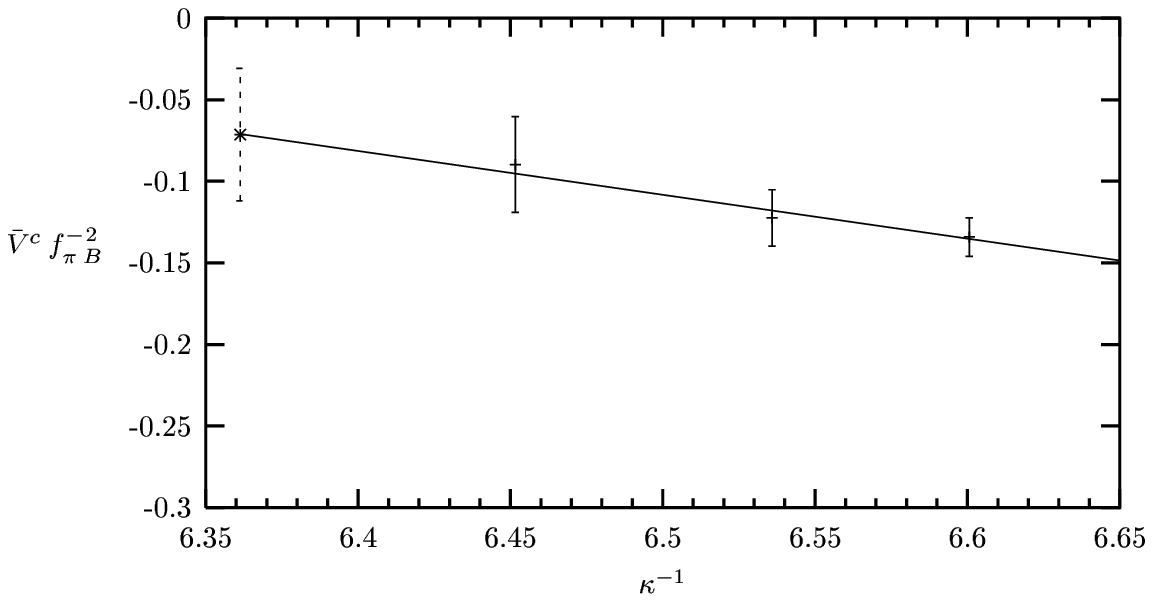, width=.85\linewidth}
\vspace*{0.3cm}

    \epsfig{file=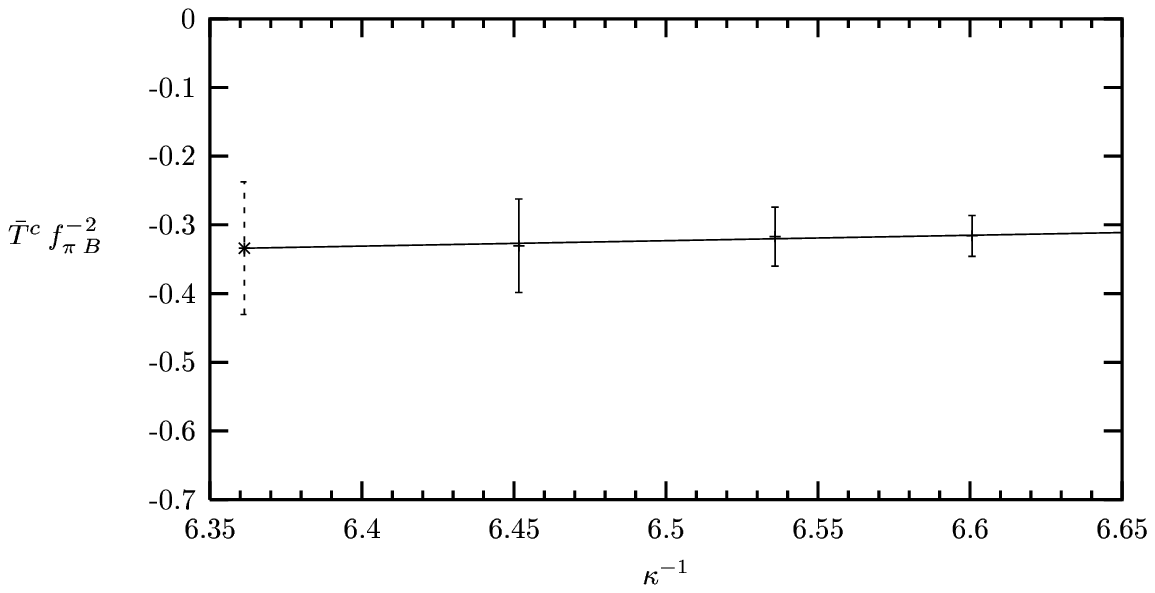, width=.85\linewidth}
\vspace*{0.3cm}
    \caption{The unrenormalized, reduced matrix elements $\bar{A}^c$,
      $\bar{V}^c$ and $\bar{T}^c$, and their extrapolations to the chiral
     limit.} 
    \label{fig:chifit2}
  \end{center}
\vspace{0.4cm}
\end{figure}

In table \ref{tab:latres} we give our results for the
unrenormalized, reduced matrix elements $\bar{O} = - ((2\kappa)^2/m_\pi)\,
\langle\pi|O_{44}|\pi\rangle^{\rm lat}$ of the various $I=2$ operators. 
The reader can
easily check that the Fierz identities (\ref{eq:fierz}) hold
identically at each value of $\kappa$. 
In fig.~\ref{fig:chifit1} we plot $\bar{O}$ for the operators without
color matrices, and in fig.~\ref{fig:chifit2} for the operators with color
matrices, as a function of $\kappa^{-1}$. The data suggest a linear
extrapolation to the chiral limit. The result of the extrapolation is
given in table~\ref{tab:latres}. The critical hopping parameter is
$\kappa_c=0.15717(3)$.

\section{Results and Conclusions}

We are now ready to give results for the structure function. To
minimize effects of higher order 
contributions to the Wilson coefficient $C_2^{(4)}$ in $g^2$, we shall take
\begin{equation}
Q^2 = \mu^2 = a^{-2}.
\end{equation}
The $g^2$ in (\ref{eq:htcoeff}) is therefore replaced by $4\pi \alpha_s(Q^2)$.
If we fix the scale by adjusting the $\rho$ mass to its physical
value, we have~\cite{Gockeler:1998fn}
$a^{-2} \approx 5 \,\mbox{GeV}^2$.
If, instead, we take the string tension or the force parameter to set
the scale, 
we have $a^{-2} \approx 4 \,\mbox{GeV}^2$.

We calculate the renormalization constants from (\ref{eq:renormres})
with $g_0=1$. Combining these with the unrenormalized lattice results
in table~\ref{tab:latres}, we find for the $I=2$ structure function
\begin{equation}
A_2^{(4)\,I=2} = 0.100(38)\, f^{\,2}_{\pi B}.
\end{equation}
Multiplying this number with the Wilson coefficient and the kinematical
factor, and expressing the result in terms of the renormalized decay
constant $f_{\pi} = Z_A f_{\pi B}$ with $Z_A = 0.867$, computed in
perturbation theory, we finally obtain
\begin{equation}
\label{final}
M_2^{I=2} =  1.67(64)\, \frac{f_{\pi}^{2}\,\alpha_s(Q^2)}{Q^2} + O(\alpha_s^2).
\end{equation}
All numbers refer to the chiral limit. 
An early calculation~\cite{Morelli:1993}, based on 15 configurations, found 
a negative value\footnote{Due to several misprints
and inconsistencies in this paper~\cite{Morelli:1993} we were not able
to trace the origin of the discrepancy.} for $M_2^{I=2}$.

It is instructive to compare (\ref{final}) with the twist-2 contribution
to the structure function of (say) the $\pi^+$.
In~\cite{Best:1997qp} we found
\begin{equation}
M_2^{(2)\,\pi^+} = 0.152(7).
\end{equation}
Relative to this number (\ref{final}) is only a small correction, except
perhaps at very small values of $Q^2$.   

In case of the nucleon we may expect similar numbers, but with $f_\pi$
being replaced by the nucleon mass. This would then result in a significant
correction. Calculations of 4-Fermi contributions to the nucleon
structure function are in progress.

\section*{Acknowledgment}
The numerical calculations have been done on the Quadrics computers at
DESY-Zeuthen. We thank the operating staff for support. This work was 
supported in part by the Deutsche Forschungsgemeinschaft.


\begin{thebibliography}{10}

\bibitem{Liuti}
S.~Liuti, Nucl. Phys. B (Proc. Suppl.) 74 (1999) 380 ({\tt hep-ph/9809248}). 

\bibitem{Gottlieb:1978}
S. Gottlieb, Nucl. Phys. B139 (1978) 125.

\bibitem{Morelli:1993}
A. Morelli, Nucl. Phys. B392 (1993) 518.

\bibitem{Capitani:1998lm}
S.~Capitani, M.~G\"ockeler, R.~Horsley, H.~Oelrich, D.~Petters,
P.~Rakow and G.~Schierholz, Nucl. Phys. B (Proc. Suppl.) 73 (1999) 288
({\tt hep-lat/9809171}).

\bibitem{Capitani:1999}
S.~Capitani, M.~G\"ockeler, R.~Horsley, D.~Petters, D.~Pleiter,
P.~Rakow, G.~Schierholz, DESY preprint DESY 99-069 ({\tt hep-ph/9906320}).

\bibitem{Jaffe:1981}
R.~L.~Jaffe and M.~Soldate, Phys. Lett. 105B (1981) 467.

\bibitem{Jaffe:1982}
R.~L.~Jaffe and M.~Soldate, Phys. Rev. D26 (1982) 49.

\bibitem{Shuryak:1982b}
E.~V. Shuryak and A.~I. Vainshtein, Nucl. Phys. B199 (1982) 451.

\bibitem{Kawai}
H.~Kawai, R.~Nakayama and K.~Seo, Nucl. Phys. B189 (1981) 40.

\bibitem{rep}
M.~G\"ockeler, R.~Horsley, E.-M.~Ilgenfritz, H.~Perlt,
P.~Rakow, G.~Schierholz and A.~Schiller, Phys. Rev. D54 (1996) 5705.

\bibitem{Okawa}
M.~Okawa, Nucl. Phys. B187 (1981) 71.

\bibitem{Rajan}
M.~Gupta, T.~Bhattacharya and S.~R.~Sharpe, Phys. Rev. D55 (1997) 4036.

\bibitem{Best:1997qp}
C.~Best, M.~G\"ockeler, R.~Horsley, E.-M.~Ilgenfritz, H.~Perlt,
P.~Rakow, A.~Sch\"afer, G.~Schierholz, A.~Schiller and S.~Schramm,
Phys. Rev. D56 (1997) 2743. 

\bibitem{Gockeler:1998fn}
M.~G\"ockeler, R.~Horsley, H.~Perlt, P.~Rakow, G.~Schierholz,
A.~Schiller and P.~Stephenson, Phys. Rev. D57 (1998) 5562.

\end{thebibliography}
\end{document}